\documentclass[twocolumn]{article}
\usepackage[a4paper, total={180mm, 243mm}]{geometry}
\usepackage[utf8]{inputenc}
\usepackage{amssymb}
\usepackage{bm}
\usepackage{soul}
\usepackage{lipsum}
\usepackage{graphicx}
\usepackage{hyperref}
\usepackage[
    backend=biber,
    style=nature,
    url=false,
    isbn=false,
    eprint=false,
    date=year,
    maxbibnames=99
]{biblatex}
\AtEveryBibitem{%
  \clearfield{note}%
}
\usepackage[T1]{fontenc}
\usepackage{siunitx}
\usepackage{upgreek}
\usepackage[version=4]{mhchem} 
\usepackage{titling}
\usepackage{circledsteps}   
\usepackage{subcaption}
\usepackage{mathtools}
\usepackage{bm}
\usepackage[inkscapeformat=pdf]{svg}

\usepackage[version=4]{mhchem}  
\usepackage[switch]{lineno} 
\modulolinenumbers[5]

\DeclareUnicodeCharacter{2009}{\,} 

\DeclareCaptionLabelSeparator{custom}{\textbar}
\DeclareCaptionFormat{custom}{\textsf{\textbf{#1 #2} \small #3}}
\newcommand\authormark[1]{\textsuperscript{#1}}

\graphicspath{{./figures}} 

\makeatletter
\newcommand{\showfontsize}{\f@size{} pt}
\makeatother

\title{Supercontinua from integrated gallium nitride waveguides}

\author{Weichen Fan\authormark{1},
        Markus Ludwig\authormark{1},
        Ian Rousseau\authormark{2},
        Ivo Arabadzhiev\authormark{2},\\
        Bastian Ruhnke\authormark{1},
        Thibault Wildi\authormark{1},
        Tobias Herr\authormark{1,3,*}}

\date{%
    \small $^1$Deutsches Elektronen-Synchrotron DESY, Notkestr. 85, 22607 Hamburg, Germany \\
    \small $^2$Hexisense SA, Rue de Genève 100, 1004 Lausanne, Switzerland\\
    \small $^3$Physics Department, Universität Hamburg UHH, Luruper Chaussee 149, 22607 Hamburg, Germany\\
    \small $^*$tobias.herr@desy.de
}

\begin{document}

\maketitle

\textbf{Supercontinua are broadband spectra that are essential to optical spectroscopy, sensing, imaging, and metrology. They are generated from ultrashort laser pulses through nonlinear frequency conversion in fibers, bulk media, and chip-integrated waveguides. For any generating platform, balancing the competing criteria of strong nonlinearity, transparency, and absence of multiphoton absorption is a key challenge.
Here, we explore supercontinuum generation in integrated gallium nitride (GaN) waveguides, which combine a high Kerr-nonlinearity, mid-infrared transparency, and a large bandgap that prevents two- and three-photon absorption in the technologically important telecom C-band, where compact erbium-based pump lasers exist. Using this type of laser, we demonstrate tunable dispersive waves and gap-free spectra extending to almost 4~$\upmu$m in wavelength, relevant to functional group chemical sensing.
Additionally, leveraging the material's second-order nonlinearity, we implement on-chip $f$-2$f$ interferometry to detect the pump laser's carrier-envelope offset frequency, which enables precision metrology. These results demonstrate the versatility of GaN-on-sapphire as a new platform for broadband nonlinear photonics.}

In supercontinuum generation, an ultrashort laser pulse creates a broadband spectrum via nonlinear optical processes \cite{dudley2006SupercontinuumGenerationPhotonic, bres2023SupercontinuumIntegratedPhotonicsa}. When the input pulses are periodic in time, the emerging spectrum can be a frequency comb, i.e., a coherent spectrum that is composed of discrete optical frequency components $\nu_m$ that are given by $\nu_m = f_\mathrm{ceo} + mf_\mathrm{rep}$, where $f_\mathrm{ceo}$ is the carrier-envelope offset frequency and $f_\mathrm{rep}$ is the repetition rate of the pulse train; $m$ is an integer line index. Frequency combs and supercontinua underpin a wide range of applications in photonics including, for instance, optical spectroscopy of molecules, environmental monitoring, frequency synthesis, optical clocks, and medical imaging \cite{picque2019FrequencyCombSpectroscopy, fortier201920YearsDevelopments, diddams2020OpticalFrequencyCombs, huang1991OpticalCoherenceTomography}. 

Complementing nonlinear fibers and bulk media, integrated nanophotonic waveguides have emerged as a powerful platform for supercontinuum generation \cite{bres2023SupercontinuumIntegratedPhotonicsa}. Dielectric and semiconductor waveguide materials including silicon \cite{kuyken2015OctavespanningMidinfraredFrequency,singh2015MidinfraredSupercontinuumGeneration}, silicon-germanium \cite{sinobad2018MidinfraredOctaveSpanninga}, (aluminum) gallium arsenide \cite{chiles2019MultifunctionalIntegratedPhotonics, kuyken2020OctavespanningCoherentSupercontinuum, granger2023GaAschipbasedMidinfraredSupercontinuum}, chalcogenides \cite{lamont2008SupercontinuumGenerationDispersiona}, silicon nitride \cite{halir2012UltrabroadbandSupercontinuumGenerationa}, aluminum nitride \cite{hickstein2017UltrabroadbandSupercontinuumGenerationa,chen2021SupercontinuumGenerationHigh}, lithium niobate \cite{lu2019OctavespanningSupercontinuumGeneration, yu2019CoherentTwooctavespanningSupercontinuum}, tantalum pentoxide \cite{jung2021TantalaKerrNonlinear, fan2019VisibleNearinfraredOctave}, tellurium oxide \cite{singh2020NonlinearSiliconPhotonics} and diamond \cite{shams-ansari2019SupercontinuumGenerationAngleetched} have enabled spectral broadening with sub-nJ pulse energies. Materials with higher nonlinearity are typically characterized by a small bandgap and are ideally pumped by long wavelength table-top laser systems, such as optical parametric oscillators, to avoid two- and three-photon absorption (2PA/3PA) and associated free-carrier-induced loss. In contrast, larger bandgap materials can be pumped at shorter near-infrared wavelength. This is illustrated for different materials in Fig.~\ref{fig_introduction}e where we also indicate the technologically important telecom C-band, where compact and robust erbium-based pump lasers are available. Large bandgap materials have therefore enabled compact setups for the generation of broadband supercontinua extending from near-infrared into ultraviolet \cite{liu2019100THzspanningUltraviolet,obrzud2019VisibleBluetored10,wu2024VisibletoultravioletFrequencyComb,cheng2024ContinuousUltravioletBluegreen,ludwig2023UltravioletAstronomicalSpectrographa} and mid-infrared domains \cite{guo2018MidinfraredFrequencyComb,lu2020UltravioletMidinfraredSupercontinuum, grassani2019MidInfraredGas, guo2020NanophotonicSupercontinuumbasedMidinfrared}.

An unexplored material for supercontinuum generation is GaN, a III-V direct bandgap semiconductor material widely used in electronics, lighting, lasers, photovoltaics, and photo-detectors \cite{nakamura1994CandelaClassHigh,mohammad1995EmergingGalliumNitride, tansu2010IIINitridePhotonics}. GaN exhibits a third-order nonlinearity $\chi^{(3)}$ larger than that of widely-used silicon nitride. The large nonlinearity is particularly appealing for high-pulse repetition rate pump sources, or for extending supercontinua to longer wavelength, where large waveguide cross-section are needed to confine the optical modes. Like lithium niobate and aluminum nitride, it also offers a high second-order nonlinearity $\chi^{(2)}$ up to 20~pm/V \cite{sanford2005MeasurementSecondOrder}. Compared to other semiconductors, its large bandgap of 3.4~eV implies the absence of 2PA and 3PA in the telecom C-band around 1550~nm, so that erbium-based pump lasers can be employed. 
Previous work on GaN has already demonstrated low-loss waveguides with transparency at near-infrared and visible wavelengths \cite{hui2003GaNbasedWaveguideDevices, xiong2011IntegratedGaNPhotonic, thubthimthong2015AsymmetricallyVerticallyCoupled, bruch2015BroadbandNanophotonicWaveguides, chen2017LowLossGaN, gromovyi2017EfficientSecondHarmonic}, efficient second harmonic generation \cite{xiong2011IntegratedGaNPhotonic,gromovyi2022LowlossGaNoninsulatorPlatform}, four-wave mixing \cite{stassen2019HighconfinementGalliumNitrideonsapphirea} and microresonator solitons \cite{zheng2022IntegratedGalliumNitride}, hinting at significant potential for supercontinuum generation.

\begin{figure*}[t!]
    \centering
    \includegraphics[width=0.9\textwidth]{fig_introduction_v11.pdf}
    \caption{(a) Photograph of a GaNOS chip pumped by a C-band femtosecond laser. (b)-(d) Colored scanning electron microscope (SEM) images of waveguide cross-section, waveguide sidewall, and microring resonator, respectively. Green marks GaN, orange marks AlN, and cyan marks sapphire. (e) Bandgap and nonlinear refractive index $n_2$ at 1560~nm of different materials. The vertical lines indicate the cut-on position of 2PA (orange) and 3PA (purple) for erbium-based lasers. Cyan and brown indicate the presence or absence of second-order nonlinearity, respectively. Refs: Diamond, AlN, \ce{Si3N4} and Si from \cite{bres2023SupercontinuumIntegratedPhotonicsa}; \ce{LiNbO3} \cite{phillips2011SupercontinuumGenerationQuasi}; \ce{Ta2O5} \cite{jung2021TantalaKerrNonlinear}; 4H-SiC \cite{guidry2020OpticalParametricOscillation}; GaN \cite{zheng2022IntegratedGalliumNitride}; \ce{As2S3} \cite{gaeta2019PhotonicchipbasedFrequencyCombs}; GaP \cite{wilson2020IntegratedGalliumPhosphide}; GaAs \cite{hurlbut2007MultiphotonAbsorptionAnd}; \ce{Al_{0.21}Ga_{0.79}As} \cite{kuyken2020OctavespanningCoherentSupercontinuum}. (f) Schematic setup. GaNOS: GaN-on-sapphire; OSA: optical spectrum analyzer; Col.: collimator; BPF: bandpass filter; PD: photodetector; ESA: electrical spectrum analyzer.  (g) \& (h) Illustration of molecular spectroscopy and $f$-$2f$ self-referencing with supercontinua.}
    \label{fig_introduction}
\end{figure*}

Here, we demonstrate for the first time the generation of broadband supercontinua in integrated GaN waveguides (Fig.~\ref{fig_introduction}a). The waveguides are fabricated via e-beam lithography and reactive ion etching from commercially available high-quality GaN-on-sapphire (GaNOS) wafers. Both GaN and sapphire are transparent deep into the mid-infrared wavelength range beyond 6~$\upmu$m. An off-the-shelf 100~MHz erbium-based mode-locked laser with a central wavelength of 1560~nm is used to pump the waveguides, resulting in the efficient generation of multi-octave spanning spectra. In tailored waveguides we demonstrate tunable dispersive wave generation and gap-free spectra extending into the mid-infrared to almost 4000~nm wavelength, ideally suited for chemical sensing and molecular spectroscopy (cf. Fig.~\ref{fig_introduction}g).
Moreover, by simultaneously leveraging second- and third-order nonlinearities, we demonstrate chip-based detection of $f_\mathrm{ceo}$ for self-referencing and optical precision metrology (cf. Fig.~\ref{fig_introduction}h).  
These results show the potential of integrated GaN waveguides for broadband nonlinear photonics, and in particular, their ability to efficiently generate mid-infrared light from erbium-based lasers without suffering from multiphoton absorption and associated free-carrier-induced loss.

\section{GaN waveguides}
The GaN waveguides used in this work are fabricated from a commercially available, unintentionally n-doped, 725~nm thick wurtzite crystalline GaN layer, grown on a sapphire substrate with a 25~nm aluminum nitride buffer layer and a defect density of $<5\times10^9$~cm$^{-2}$. The c-axis of the crystalline GaN is perpendicular to the wafer surface. In this crystal orientation, the transverse magnetic (TM) polarized modes experience the largest second-order nonlinear susceptibility $\chi_\mathrm{33}^{(2)}=$ 10-20~pm/V \cite{sanford2005MeasurementSecondOrder}.
The waveguides are defined via e-beam lithography with hydrogen silsesquioxane (HSQ) resist, followed by inductively-coupled plasma reactive ion etching with a chlorine/nitrogen mixture \cite{rousseau2017QuantificationScatteringLoss}. A complete etching through the GaN and aluminum nitride layers results in waveguides with well-defined, symmetric sidewall angles of \SI{73}{\degree}. Finally, the HSQ is removed by hydrofluoric acid, and the chips are cleaved to create the input and output coupling facets. The root mean square roughness of the top surface is less than 3~nm after resist removal, measured in a 5${\times}$\SI{5}{\micro m^2} area. Figure~\ref{fig_introduction}b shows a typical waveguide cross-section (after cleaving), and Fig.~\ref{fig_introduction}c provides a lateral view of the low sidewall roughness waveguides.
Waveguides with two different lengths, 2~mm and 5~mm, are fabricated.

\begin{figure*}[t!]
    \centering
    \includegraphics[width=0.9\textwidth]{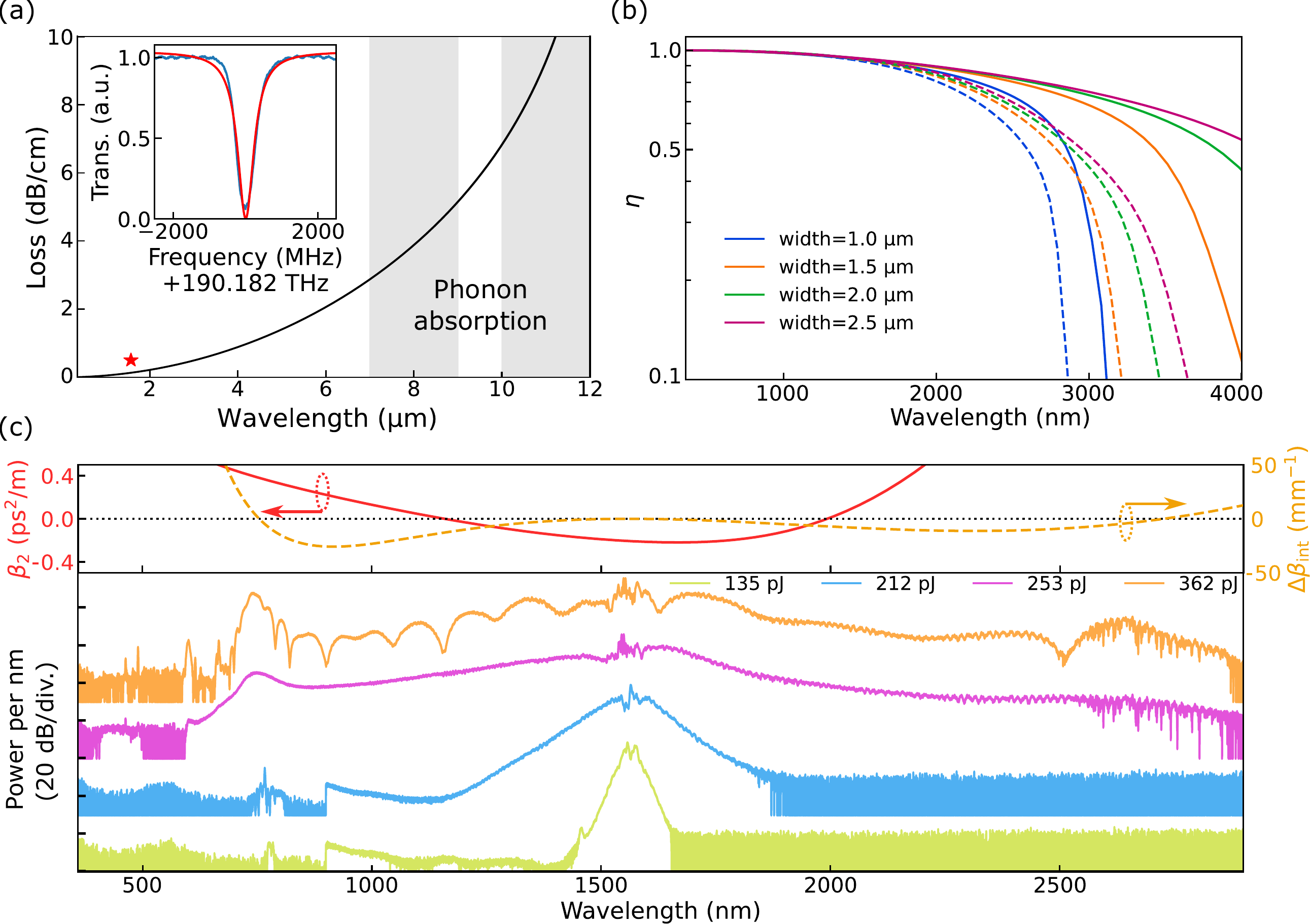}
    \caption{(a) Estimated material loss versus wavelength with unintentionally doped GaN using the Drude model. Grey shaded areas indicate phonon absorption bands and the red star marks the propagation loss measured by a microring. Inset: resonance of the microring (blue) and Lorentzian fit (red) with a full-width-at-half-maximum linewidth of 543.50~MHz. (b) Simulated mode confinement versus wavelength for different waveguide widths. Solid lines are fundamental TE modes, and dashed lines are fundamental TM modes. (c) Upper panel: group velocity dispersion (solid red) and integrated dispersion (dashed orange) for the TM polarized mode in a 2.5~$\upmu$m wide waveguide. Lower panel: supercontinua obtained from a corresponding waveguide of 2~mm length; for visibility the spectra are vertically offset by 30~dB.}
    \label{fig_characterization}
\end{figure*}

To determine the waveguide loss, we measure the Q-factor (linewidth) of a microring-resonator with a waveguide width of 2.5~$\upmu$m and a radius of 39~$\upmu$m (Fig.~\ref{fig_introduction}d). Under critical coupling, we observe a loaded Q-factor of 3.5$\times$10$^{5}$ (linewidth 543.50~MHz) at the wavelength of 1576~nm (Fig.~\ref{fig_characterization}a, inset), corresponding to a waveguide propagation loss of 0.53~dB/cm.
In semiconductor materials, a significant contribution to the propagation loss can arise from free-carriers. In a Drude-model~\cite{kasic2000FreecarrierPhononProperties} these losses scale with $\lambda^2$, where $\lambda$ is the wavelength (see details in the Supplementary Information, SI).
As Hall effect measurements reveal, the unintentional n-doping in our samples leads to a free-carrier concentration of 2.1$\times$10$^{16}$~cm$^{-3}$.
Based on this measurement and assuming an electron mobility of 800~cm$^2/$V \cite{kaess2016CorrelationMobilityCollapse}, we can use the Drude model to estimate the wavelength-dependent contribution of the unintentional n-doping to the propagation loss.
At a wavelength of 1576~nm, we find the free-carrier contribution to the loss to be 0.14~dB/cm. As indicated in Fig.~\ref{fig_characterization}a, this is consistent with the microring measurement, which includes other loss mechanisms, such as scattering. While the unintentional n-doping should negligibly affect the waveguide losses at shorter wavelengths (as long as direct absorption in GaN, 2PA, and 3PA can be neglected), an increasing impact towards longer wavelengths is expected based on the Drude model, approaching 3~dB/cm at 7~$\upmu$m. For even longer wavelengths, first and second harmonic phonon absorption bands (see absorption regions shaded in gray in Fig.~\ref{fig_characterization}a) \cite{yang2005PhotonAbsorptionRestrahlen, welna2012TransparencyGaNSubstrates} would likely become the predominant source of loss. Moreover, beyond a wavelength of $\sim$6~$\upmu$m, losses in the sapphire substrate can no longer be neglected.

In addition to low loss propagation, tight mode confinement in the waveguide core is critical to enable efficient and broadband supercontinuum generation. To characterize the wavelength- and polarization-dependent mode confinement, we perform numerical simulations based on a finite element model (FEM). The model includes previously obtained material data of GaN \cite{bowman2014BroadbandMeasurementsRefractive} while the refractive index of sapphire and aluminum nitride are measured by ellipsometry. 
Figure~\ref{fig_characterization}b shows the mode-confinement defined by
\begin{equation}
    \eta=\frac{\iint_{GaN} I(x,y) \,dx\,dy}{\iint_{all} I(x,y) \,dx\,dy},
\end{equation} 
for both, fundamental transverse electric (TE) and TM modes, where $I(x,y)$ is the intensity profile across the waveguide's cross-section. For wide waveguide, a mode-confinement of more than 50\% of the power in the GaN core can be maintained up to a wavelength of 4000~nm in the TE polarization. In contrast, the TM mode confinement is much smaller at longer wavelengths. 
To achieve efficient coupling to the chip via lensed fibers, the waveguides are tapered to a width of 4.0~$\upmu$m at the coupling facets for 5~mm long chips. In the experiment, we observe a per-facet coupling loss of 4.6~dB with polarization maintaining (PM) lensed fiber with a spot size of \SI{2.5}{\micro m}. While the symmetry of the lensed fiber mode suppresses coupling to antisymmetric higher-order modes, we cannot exclude that a small fraction of the input light is projected into symmetric higher-order modes. This is not expected to impact supercontinuum generation, however might lead to an overestimation of the relevant on-chip pulse energies.

\section{Supercontinuum generation}
To generate supercontinua, we utilize an off-the-shelf erbium-doped mode-locked laser with 100~MHz pulse repetition rate (Fig.~\ref{fig_introduction}f) and a center wavelength of 1560~nm. The laser is amplified to a maximum of 150~mW of average power in an erbium-doped fiber amplifier, and a waveshaper is included in the setup prior to amplification for dispersion compensation, resulting in pulses as short as 55~fs. A lensed fiber couples the pulses to the chip. All optical components before the chip are PM and single-mode.
The generated light is collected by a multi-mode fluoride fiber via butt-coupling and measured by optical spectrum analyzers (OSAs).\\
Figure~\ref{fig_characterization}c shows broadband supercontinua that are observed in TM polarization with different pulse energies in a only 2~mm long waveguide of 2.5~$\upmu$m width (nonlinear coefficient $\gamma$=3.23~m$^{-1}$W$^{-1}$). 
Spectra in excess of two octaves are obtained with 253~pJ of pulse energy, wherein the 30~dB bandwidth exceeds 1.5 octaves. A pronounced dispersive wave (DW) is apparent at approximately half the pump wavelength; expected from the zero-crossing of the integrated dispersion $\Delta\beta_\mathrm{int}(\omega)=\beta(\omega)-\beta(\omega_0)-\beta_{1,\omega_0}(\omega-\omega_0)$, where $\omega$ is the frequency, $\beta$ is the wavenumber, $\omega_0$ is the central frequency of the pump laser and $\beta_1=\mathrm{d}\beta/\mathrm{d}\omega$. The obtained spectra are dispersion limited; increasing the pump power further or utilizing longer waveguides would not result in a significant growth in span.As an aside, we note that in our experiments, we have not observed  damage or degradation of the waveguides, including when testing them with a continuous-wave laser at 1560\,nm with 1\,W of incident power.

\begin{figure*}[t!]
    \centering
    \includegraphics[width=0.9\textwidth]{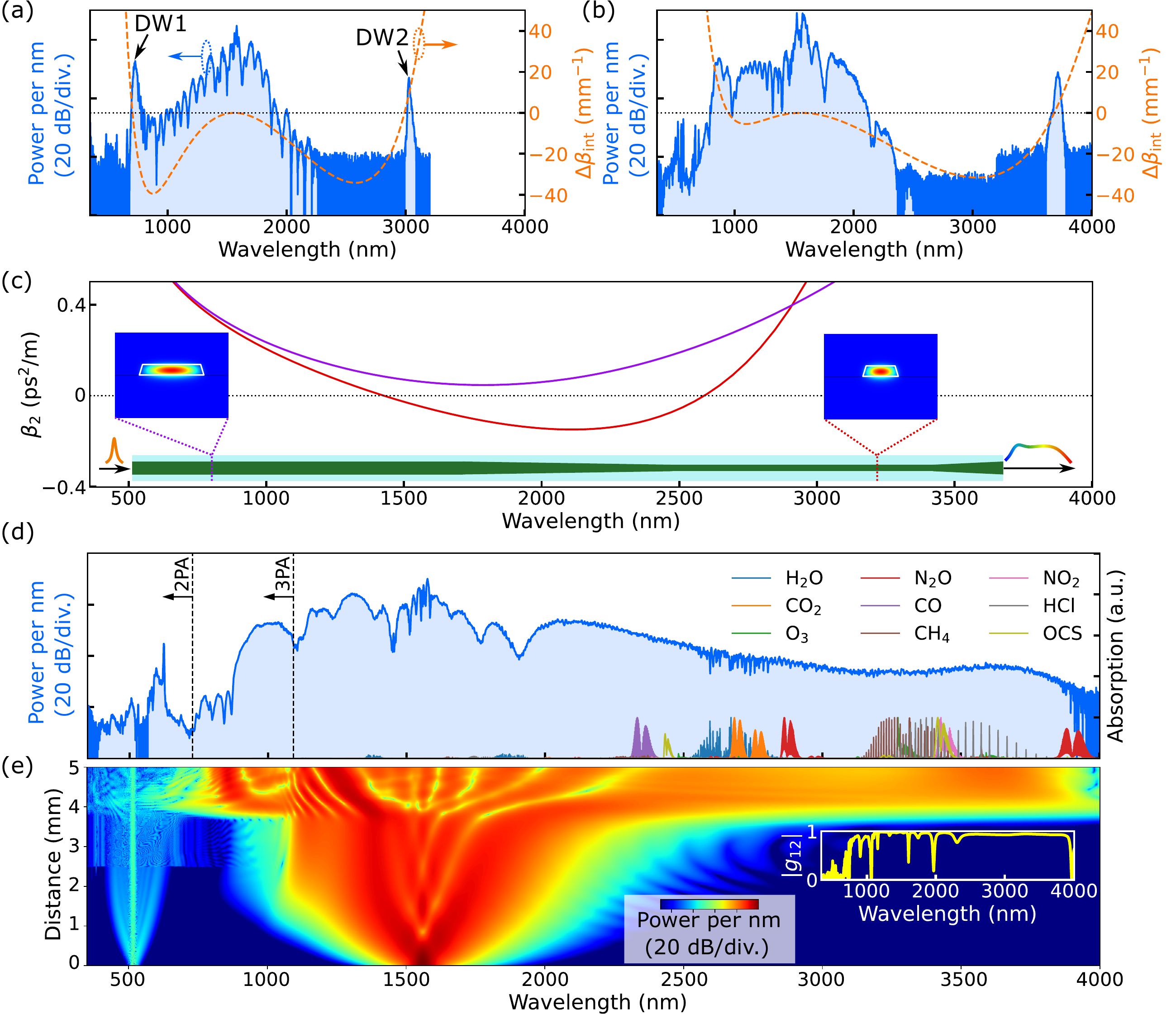}
    \caption{(a) \& (b) Supercontinua obtained from 2~mm long waveguides (blue), and simulated integrated dispersion (dashed orange lines) of fundamental TE mode for waveguide widths of 1.0~$\upmu$m and 1.5~$\upmu$m, respectively. (c) Simulated group velocity dispersion of 4.0~$\upmu$m (purple) and 2.0~$\upmu$m (red) wide waveguides. The inset shows the waveguide's width profile (top view) and the simulated mode fields at the pump wavelength in both segments of the waveguides. 
    (d) Experimental spectrum obtained from a 5~mm long segmented waveguide (blue). The overlay represents normalized absorption spectra of different gases \cite{kochanov2016HITRANApplicationProgramming}. (e) Simulated spectral evolution \cite{voumard2023SimulatingSupercontinuaMixed} and the inset is calculated first order mutual coherence $|g_{12}|$.}
    \label{fig_TE}
\end{figure*}

\subsection{Mid-infrared supercontinua}
Mid-infrared spectra in the functional group fingerprint window from 2-5~$\upmu$m are of high relevance to chemical and molecular sensing \cite{schliesser2012MidinfraredFrequencyCombs}.
Therefore, we explore the platform's ability to support the generation of mid-infrared light from an erbium-based pump laser.
We now operate the waveguides in TE polarization to benefit from the stronger mode confinement at longer wavelengths (Fig.~\ref{fig_characterization}b). Two DW positions are predicted by the integrated dispersion of these waveguides, where the longer wavelength one falls into the mid-infrared and can be tuned from around 3000~nm to $>$3700~nm by adjusting the waveguide width from 1.0~$\upmu$m to 1.5~$\upmu$m, similar to observations in silicon nitride and aluminum nitride waveguides \cite{guo2018MidinfraredFrequencyComb,grassani2019MidInfraredGas,lu2020UltravioletMidinfraredSupercontinuum}. Figures~\ref{fig_TE}a and b show the experimental results obtained in 2~mm long waveguides, accurately matching our prediction, despite the large spectral separation from the pump.
We attempted shifting the DW to even longer wavelengths (using a 1.8~$\upmu$m wide waveguide); however, we did not observe any DW at the expected wavelength of 3900~nm. We attribute the absence of a longer wavelength DW to the weak mode-confinement of the only 725~nm thick waveguide layer and vanishing anomalous dispersion at the pump wavelength (preventing short pulse soliton dynamics that drive the DW).

In a separate experiment, we aim at generating a gap-less, unstructured mid-infrared spectrum as needed for broadband spectroscopy, e.g. multi-molecular species detection. To achieve this, we design a segmented waveguide (Fig.~\ref{fig_TE}c, inset) comprising a 1.5~mm long normal group velocity dispersion segment (4.0~$\upmu$m waveguide width), followed by 1.2~mm long tapered waveguide that connects to a 1.5~mm long segment with anomalous group velocity dispersion between 1500~nm and 2600~nm (2.0~$\upmu$m waveguide width). 
Such a design first broadens the pump in the normal dispersion part, which is then compressed to ultrashort temporal duration (and broad spectral bandwidth) in the anomalous dispersion part. Experimentally, a gap-free spectrum ranging approximately from 870~nm to 3900~nm is generated in the 5~mm long (including coupling tapers) segmented waveguide driven by 527~pJ on-chip energy with 55~fs pulse duration (Fig.~\ref{fig_TE}d), much broader than the spectra from waveguides of the same length but constant width (see Fig.~S2 in the SI). Numerical simulations via \textit{pyChi} \cite{voumard2023SimulatingSupercontinuaMixed} (Fig.~\ref{fig_TE}e, $\chi^{(2)}$ set to zero, see SI for details) agree well with the experimental results, and suggest that the broadband spectrum is associated with an ultrashort temporal feature of $<$10~fs duration obtained during compression. The smooth and gap-free spectrum covers a large portion of the 2-5~$\upmu$m sensing window.
The first order mutual coherence $|g_{12}|$ \cite{dudley2002CoherencePropertiesSupercontinuum} is calculated based on 200 pairs of spectra (400 independent simulations) with random shot noise included in the initial pump pulse, revealing high coherence as shown in Fig.~\ref{fig_TE}e, inset.
The strong spectral absorption features in the spectrum between 2500~nm and 2900~nm are due to atmospheric water absorption, which occurs mostly inside the optical spectrum analyzer and to smaller amount on the chip. Importantly, as we illustrate in Fig.~\ref{fig_TE}d, the generated spectrum covers the absorption spectra of important molecular species, including those of the major greenhouse gases CO$_2$, N$_2$O, and CH$_4$. In contrast to platforms utilizing fused silica cladding, we do not expect material loss to be responsible for the spectrum's long-wavelength cutoff, and attribute it to reduced mode confinement and related strong normal dispersion at long wavelengths. These limitations result from the thin GaN layer and are not fundamental; they may be overcome through increased layer thickness and optimized waveguide geometries in future work. Regarding long-wavelength generation, we highlight the large bandgap of GaN, which effectively suppresses free-carrier generation through 2PA and 3PA at the pump wavelength, and of 2PA across nearly the entire spectrum. This prevents free-carrier generation and unwanted impact on the generated spectrum \cite{kuyken2020OctavespanningCoherentSupercontinuum}, particularly in the mid-infrared, due to the $\lambda^2$-scaling of free-carrier induced loss.

\begin{figure*}[t!]
    \centering
    \includegraphics[width=0.9\textwidth]{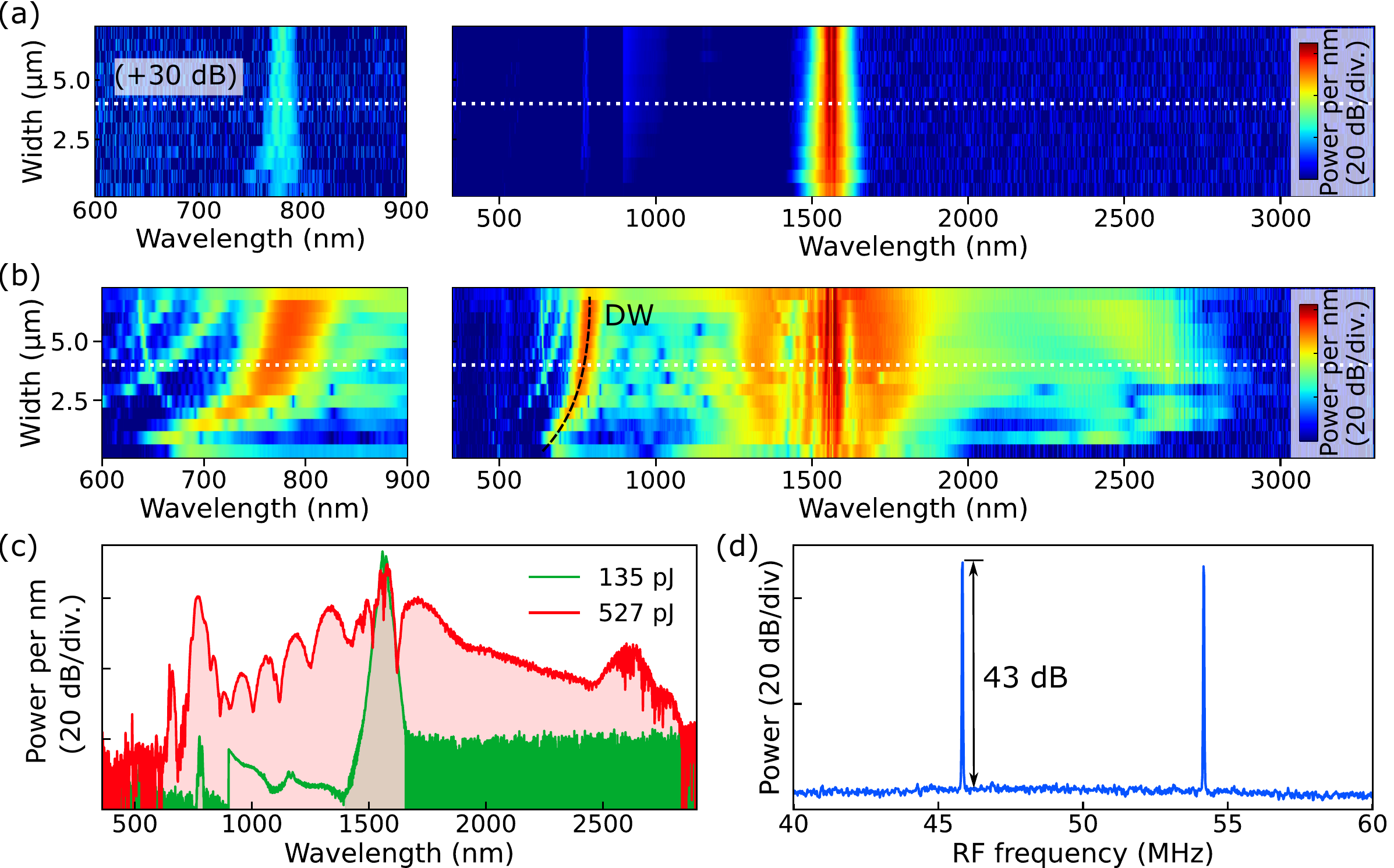}
    \caption{(a) \& (b) Measured spectra for different waveguide widths with $\sim$135~pJ and $\sim$527~pJ on-chip pump pulse energy. The left panels show a magnified view of the right panels between 600~nm and 900~nm. (c) Output spectra from a 4.0~$\upmu$m wide waveguide, corresponding to slices marked by dotted white lines in (a) and (b). (d) Measured $f_\mathrm{ceo}$ beanotes after bandpass-filtering a 10~nm wide portion around 780~nm of the spectrum shown in (c). The beatnote signal is recorded with a 10~kHz resolution bandwidth (RBW) and 100~Hz video bandwidth.}
    \label{fig_beatnote}
\end{figure*}

\subsection{On-chip f-2f interferometry}
The ability to efficiently generate (multi-)octave spectra is an important prerequisite for self-referencing of optical frequency combs, where $f_\mathrm{ceo}$ of the pulse train is measured via $f$-$2f$ interferometry. In this scheme, a longer wavelength portion of the spectrum is frequency-doubled (sum-frequency generation) to overlap with the original spectrum at a shorter wavelength. Photodetection of the beating between the doubled and original spectrum yields $f_\mathrm{ceo}$. As GaN not only offers a strong third-order nonlinearity for broadband spectrum generation but also possesses a second-order nonlinearity, this enables implementation of the entire $f$-$2f$ interferometer in a GaN waveguide, similar to approaches based on lithium niobate \cite{yu2019CoherentTwooctavespanningSupercontinuum,okawachi2020ChipbasedSelfreferencingUsing,  obrzud2021StableCompactRFtoopticala} and aluminum nitride \cite{hickstein2017UltrabroadbandSupercontinuumGenerationa}, as well as, self-organized gratings \cite{hickstein2019SelforganizedNonlinearGratings} and third harmonic generation \cite{carlson2017SelfreferencedFrequencyCombs} in silicon nitride. To access the largest $\chi^{(2)}$-tensor element of GaN for frequency doubling, we pump the waveguides in the TM polarization. Initially, to validate the frequency doubling, we pump the 2~mm long waveguides with a pulse energy of 135~pJ, low enough to isolate a second harmonic signal from the broadband supercontinuum. In all tested waveguides with waveguide widths ranging from 0.4~$\upmu$m to 7.0~$\upmu$m, we observe a second harmonic signal of the pump at 780~nm (Fig.~\ref{fig_beatnote}a), which is limited in power and bandwidth by the non-zero phase mismatch. To generate a broadband supercontinuum, reaching 780~nm, we increase the pulse energy to 527~pJ. As expected based on the integrated dispersion, the spectra exhibit a pronounced DW ranging from approximately 700~nm to 800~nm, depending on waveguide width (Fig.~\ref{fig_beatnote}b). To achieve a high signal to noise ratio (SNR), we use a waveguide width of 4.0~$\upmu$m, as here the spectral position of the DW is well matched with the frequency-doubled pump laser (Fig.~\ref{fig_beatnote}c).
Differing from the experiments where we record optical spectra, here for $f_\mathrm{ceo}$ detection we collect the light from the chip using a lensed fiber (multi-mode at 780~nm) and collimate it to free space for spectral filtering with a 10~nm wide bandpass filter centered at 780~nm. The light is focused on a silicon photodiode, and the resulting radio-frequency signal is measured by an electrical spectrum analyzer (ESA) (Fig.~\ref{fig_introduction}f).
The observed $f_\mathrm{ceo}$ signal is shown in Fig.~\ref{fig_beatnote}d, exhibiting a SNR of 43~dB in a resolution bandwidth (RBW) of 10~kHz. This is more than sufficient for phase-locking of $f_\mathrm{ceo}$ in a self-referenced frequency comb for optical precision metrology. 
Further improvements of the SNR would likely be possible by an optimized and single-mode output coupling for the 780~nm spectral portion.

\section{Discussion}
In summary, we demonstrate efficient generation of ultra-broadband, multi-octave supercontinua in GaN-on-sapphire waveguides pumped by an off-the-shelf C-band erbium-based femtosecond laser. We showcase two important use cases of supercontinua, $f$-2$f$ interferometry for carrier-envelope offset frequency detection in a waveguide, and efficient mid-infrared supercontinuum generation extending to nearly 4~$\upmu$m that can support molecular spectroscopy. 
These results highlight the potential of GaN-on-sapphire for advancing broadband nonlinear photonics, and extending erbium-based laser technology to the mid-infrared wavelength range without limitations from two- and three-photon absorption (and associated free-carriers). 
The platform promises to extend supercontinua to even longer wavelength, where thicker GaN layers can maintain tight mode confinement. Potential free-carrier losses from unintentionally doped materials might become relevant at long wavelengths, but could be mitigated by semi-insulating GaN \cite{tanaka2023ComparativeStudyElectrical}.  
Cladding the chips e.g. with \ce{Al2O3}, AlN or \ce{MgF_2} can avoid atmospheric absorption and may provide additional opportunity for optimized inverse taper input coupling, dispersion engineering and reduced propagation loss. Moreover, waveguides implemented in the AlN-GaN material system, could enable supercontinua even beyond the transparency window of sapphire substrates.
Finally, fabricating waveguides from periodically orientation-poled GaN films \cite{pezzagna2005SubmicronPeriodicPoling,hite2012DevelopmentPeriodicallyOriented}, could lead to opportunities for chip-integrated quasi-phase-matching in second-order nonlinear processes.

\subsection*{Funding}
This project has received funding from the European Research Council (ERC, grant agreement No 853564), the European Innovation Council (EIC, grant agreement No 101046920), and through the Helmholtz Young Investigators Group VH-NG-1404.

\subsection*{Acknowledgement}
The work was supported through the Maxwell computational resources operated at DESY. We thank Yokogawa Deutschland GmbH for lending us a mid-infrared optical spectrum analyzer (AQ6377). We thank the Center for Hybrid Nanostructures (CHyN), Universität Hamburg, for offering access to the scanning electron microscope. We also thank Nicolas Grandjean and Raphaël Butté for valuable discussions.

\subsection*{Data availability}
Data underlying the results presented in this paper are available from the corresponding author upon reasonable request.

\subsection*{Disclosure}
The authors declare no conflicts of interest. I.R. is founder of Hexisense.

\printbibliography

\clearpage

\onecolumn

\title{Supplementary information to: Supercontinua from integrated gallium nitride waveguides}
\author{}
\date{}
\maketitle

\setcounter{section}{0}
\setcounter{figure}{0}
\setcounter{equation}{0}
\renewcommand{\theequation}{S\arabic{equation}}
\renewcommand{\thefigure}{S\arabic{figure}}

\section{Derivation of the free-carrier induced loss based on the Drude model}

In the following we give detailed steps to derive the free-carrier induced loss from the well-known Drude model \cite{kasic2000FreecarrierPhononProperties}. Since the GaN in this work is unintentionally n-type doped, we only take electrons into consideration. Based on the Drude model, the complex permittivity $\epsilon$ including free-carrier effects is:

\begin{equation}
    \epsilon = \epsilon'_r+\mathrm{i}\epsilon'_i = \epsilon_r+\mathrm{i}\epsilon_i-\frac{\omega_p^2}{\omega^2+\mathrm{i}\omega\gamma_p},  
\end{equation}
\begin{equation}
    \omega_p^2 = \frac{N_ee^2}{\epsilon_0m_e^{*}},
\end{equation}
\begin{equation}
    \gamma_p = \frac{e}{m_e^{*}\mu_e},
\end{equation}
where $\epsilon_r+\mathrm{i}\epsilon_i$ is the intrinsic material permittivity, $\omega_p$ and $\gamma_p$ are the plasma and damping frequency of the free-carriers, respectively, $N_e$ is the free-carrier concentration, $e$ is the electron charge, $\epsilon_0$ is the vacuum permittivity, $m_e^{*}$ is the effective electron mass, and $\mu_e$ is the electron mobility. We write the complex refractive index $\tilde{n}$ as:
\begin{equation}
    \tilde{n} = n'+\mathrm{i}k' = n+n_\mathrm{fc}+\mathrm{i}(k+k_\mathrm{fc}),
\end{equation}
\begin{equation}
    \epsilon = \tilde{n}^2, 
\end{equation}
where $n_\mathrm{fc}+\mathrm{i}k_\mathrm{fc}$ is the free-carrier induced change of refractive index. Combining Eqs.~(S1), (S4) and (S5) yields:
\begin{equation}
    {n'}^2 = \frac{\epsilon'_r+\sqrt{{\epsilon'_r}^2+{\epsilon'_i}^2}}{2}.
\end{equation}
\begin{equation}
    {k'}^2 = \frac{-\epsilon'_r+\sqrt{{\epsilon'_r}^2+{\epsilon'_i}^2}}{2}.
\end{equation}
In low loss materials $\epsilon'_i \ll \epsilon'_r$, thus the complex refractive index can be simplified to:
\begin{equation}
    n' = \sqrt{\epsilon'_r},
\end{equation}
\begin{equation}
    k' = \frac{\epsilon'_i}{2\sqrt{\epsilon'_r}}.
\end{equation}
Combining Eqs.~(S1), (S4), (S7) \& (S8) and neglecting the contribution of $n_\mathrm{fc}$ to the real part of refractive index, one can obtain:
\begin{equation}
    k_\mathrm{fc} = \frac{1}{2n}\frac{\omega_p^2\gamma_p}{\omega^3+\omega\gamma_p^2} \approx \frac{\omega_p^2\gamma_p}{2n\omega^3}.
\end{equation}
Substituting Eqs.~(S2) and (S3) into Eq.~(S10) results in:
\begin{equation}
    k_\mathrm{fc} = \frac{e^3\lambda^3}{16\pi^3c^3\epsilon_0n}\frac{N_e}{{m_e^{*}}^2\mu_e}.
\end{equation}
Then the free-carrier induced loss is, as shown previously \cite{soltani2015FreeCarrierElectrorefraction}:
\begin{equation}
    \alpha_\mathrm{fc} = \frac{4{\pi}k_\mathrm{fc}}{\lambda} = \frac{e^3\lambda^2}{4\pi^2c^3\epsilon_0n}\frac{N_e}{{m_e^{*}}^2\mu_e}.
\end{equation}

\section{Supercontinuum from the segmented waveguide, supplementary figure.}

\begin{figure*}[ht!]
  \centering
  \includegraphics[width=0.9\textwidth]{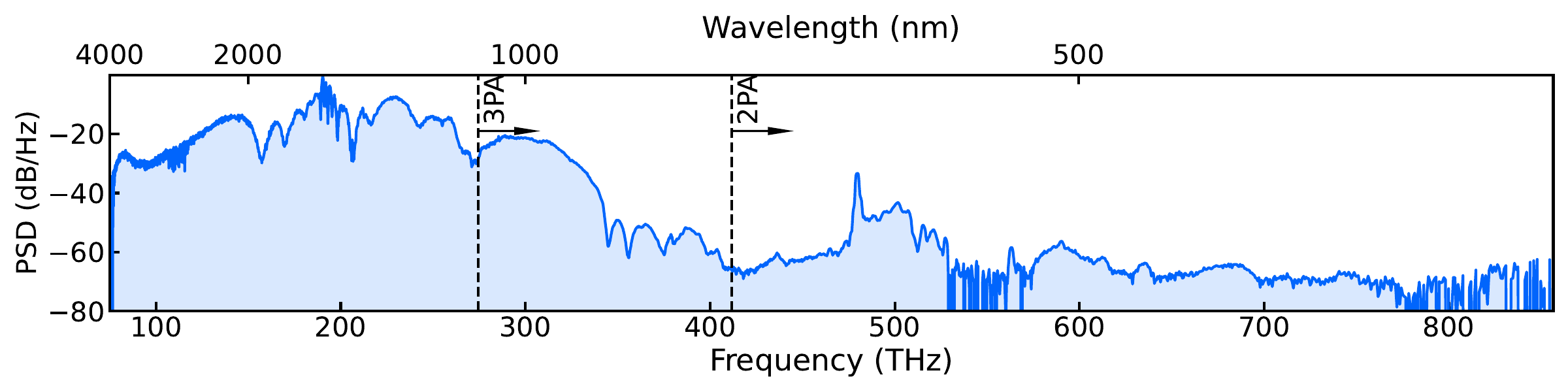}
  \caption{Power spectral density of the supercontinuum shown in the main manuscript in Fig.~3d.}
  \label{frequency_plot}
\end{figure*}

Figure~\ref{frequency_plot} shows the power spectral density (PSD) of the supercontinuum generated in the segmented waveguide (Fig.~3d in the main manuscript) plotted as a function of frequency.

\section{Comparison between segmented and straight waveguides}

\begin{figure*}[ht!]
  \centering
  \includegraphics[width=0.9\textwidth]{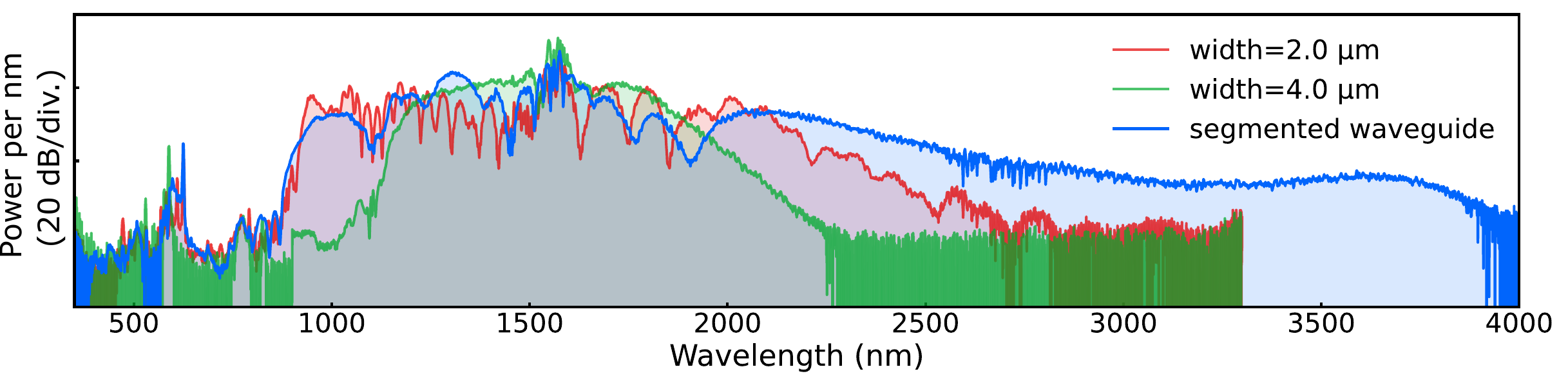}
  \caption{Comparison of supercontinua from the segmented waveguide (blue), \SI{4}{\micro m} wide waveguide (green), and \SI{2}{\micro m} wide waveguide (red). All waveguide lengths are 5~mm.}
  \label{spec_comparison}
\end{figure*}

Figure~\ref{spec_comparison} compares the spectrum from the segmented waveguide (Fig.~3d in the main manuscript) with the results from 5~mm long \SI{4}{\micro m} and \SI{2}{\micro m} wide waveguides. It shows that the dispersion tailoring in segmented waveguides leads to a broader bandwidth, and in particular facilitates the generation of mid-infrared light.

\section{Role of $\chi^{(2)}$ in the simulation of the segmented waveguide supercontinuum}

\begin{figure*}[ht!]
  \centering
  \includegraphics[width=0.9\textwidth]{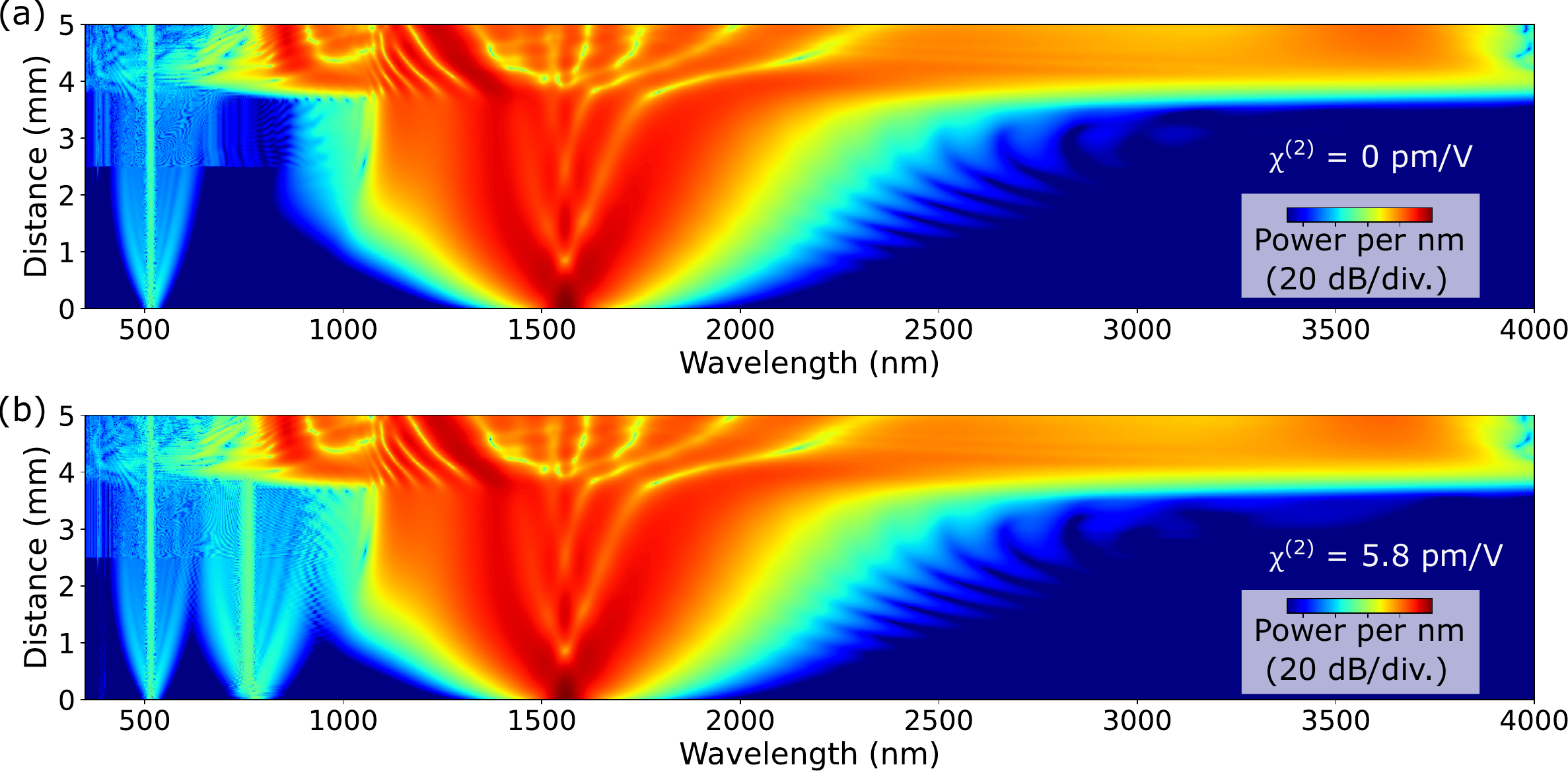}
  \caption{Simulated spectral evolution of the segmented waveguide with (a) $\chi^{(2)}=0$~pm/V and (b) $\chi^{(2)}=5.8$~pm/V.}
  \label{compare_with_without_chi2}
\end{figure*}

In the experiment shown in manuscript in Fig.~3, the segmented waveguide is driven in the TE polarization,perpendicular to the c-axis of GaN. This implies that the pump light experiences the second-order susceptibility $\chi_{31}^{(2)}$. Light generated through $\chi_{31}^{(2)}$ would be in the TM polarization.
As from a physics perspective phase matching between the orthogonal fundamental modes is not expected according to simulated waveguide effective refractive indices, and from a technical perspective, our simulation code \textit{pyChi} is currently not capable of simulating two different polarization modes, we have omitted second order nonlinear effects (i.e. $\chi^{(2)}=0$~pm/V) from the simulation shown in Fig.~3e in the main manuscript. 
To estimate a potential impact of the second order nonlinearity, we perform an additional simulation where we set the scalar $\chi^{(2)}=5.8$~pm/V with the refractive index data of the fundamental TE mode. This should provide a good estimate. As the comparison between both simulations in Fig.~\ref{compare_with_without_chi2} reveals, the impact of the second order nonlinear effects is not noticeable.

\end{document}